\documentclass[twocolumn,aps,prc,showpacs,showkeys,superscriptaddress,footinbib,eqsecnum]{revtex4-1}

\usepackage{graphicx}
\usepackage{amsmath}
\usepackage{amsfonts}
\usepackage{amssymb}
\usepackage{bm}
\usepackage{subfigure}
\usepackage{hyperref}

\begin{document}

\title{Single top quark production in heavy ion collisions at the LHC}
\author{A.V. Baskakov}
\affiliation{ Skobeltsyn Institute of Nuclear Physics, Lomonosov Moscow State University, Moscow, Russia}
\author{E.E. Boos}
\affiliation{ Skobeltsyn Institute of Nuclear Physics, Lomonosov Moscow State University, Moscow, Russia}
\author{L.V. Dudko}
\affiliation{ Skobeltsyn Institute of Nuclear Physics, Lomonosov Moscow State University, Moscow, Russia}
\author{I.P. Lokhtin}
\affiliation{ Skobeltsyn Institute of Nuclear Physics, Lomonosov Moscow State University, Moscow, Russia}
\author{A.M. Snigirev}
\affiliation{ Skobeltsyn Institute of Nuclear Physics, Lomonosov Moscow State University, Moscow, Russia}


\begin{abstract}
The paper presents analysis of the single top quark production in PbPb collisions at the Large Hadron
Collider at center-of-mass energy 5.5 TeV per nucleon pair. The analysis is performed with CompHEP and 
PYQUEN event generators. The neutron and proton content in the nuclei is taken into account. 
NLO precision has been implemented to simulate kinematic properties and rate of single top production. The modification of different characteristics of single top quark decay products due to interactions of jet partons in quark-gluon medium, and the specific charge asymmetry of
top/anti-top quark yields due to the isospin effect are evaluated.
\end{abstract}

\maketitle

\section{Introduction}

One of the important tools to study the properties of hot and dense matter created in
ultrarelativistic heavy ion collisions is a QCD jet production. Medium-induced
energy loss of energetic partons, the so-called jet quenching, is supposed to
be very different in cold nuclear matter and in quark-gluon medium, and leads to a number of 
phenomena which have been already seen in the RHIC data on high momentum particle production in 
gold-gold collisions (see, e.g., 
reviews~\cite{d'Enterria:2009am,Wiedemann:2009sh,Accardi:2009qv,Majumder:2010qh,Dremin:2010jx} 
and references therein). The LHC heavy ion program makes it possible to probe the new 
frontiers of the high temperature Quantum Chromodynamics (QCD) with the increasing role of hard 
and massive particle production processes. The effective reconstruction of energetic jets and 
other hard probes (like W-boson and b-tagged jets) at LHC energies becomes possible even in 
heavy ion environment. A number of interesting LHC results from PbPb runs at 
$\sqrt s_{\rm NN}=2.76$ TeV, have been published by ALICE, ATLAS and CMS collaborations 
(the overview of the results from the first year of heavy ion physics at LHC can be found 
in~\cite{Muller:2012zq}). In particular, the first direct observation of jet quenching has been 
done measuring the transverse energy imbalance in dijet~\cite{Aad:2010bu,Chatrchyan:2011sx,Chatrchyan:2012nia} and 
photon+jet~\cite{Chatrchyan:2012gt} channels. Then jet quenching effect has been manifested 
by a number of specific phenomena, such as the suppression of 
inclusive~\cite{Aad:2012vca,Abelev:2013kqa,Aad:2014bxa} and semi-inclusive~\cite{Adam:2015doa} 
jet rates in central PbPb collisions compared to peripheral events and proton-proton interactions, 
similar suppression for jets from b-quark fragmentation~\cite{Chatrchyan:2013exa}, azimuthal 
anisotropy of jet spectra~\cite{Aad:2013sla}, and medium-modifed jet internal 
structure~\cite{Chatrchyan:2014ava,Aad:2014wha} (see~\cite{Spousta:2013aaa} for some overview). 
At that a number of theoretical calculations and Monte-Carlo simulations in many different approaches 
were attempted to reproduce the jet quenching observables in PbPb collisions at the LHC, and to 
extract by such a way new information about created medium and partonic energy loss mechanisms~\cite{CasalderreySolana:2010eh,Qin:2010mn,Young:2011qx,Srivastava:2011nq,Lokhtin:2011qq,Betz:2012qq,Renk:2012cx,Renk:2012cb,Renk:2013rla,Apolinario:2012cg,Zapp:2012ak,Kharzeev:2012re,Dai:2012am,Huang:2013vaa,Zakharov:2012fp,Zakharov:2013gya,Burke:2013yra,Xu:2014ica,Mehtar-Tani:2014yea,Casalderrey-Solana:2014bpa,Perez-Ramos:2014mna,Lokhtin:2014vda}. The review of the current
status of jet quenching theory can be found e.g. in~\cite{Majumder:2014vpa}.

Single top quark production is a new interesting hard probe of quark gluon matter at the LHC.
Top quarks are produced and decay at very early stage of the nuclear reaction before the dense 
matter formation. Then partonic products of top quark decay loose the energy propagating through 
the QCD-medium, and so their different kinematic distributions get modified. In particular, 
smearing and decreasing mean and maximum values of invariant mass distributions of 
three jets from top quark decay (one b-jet and two jets from W-boson) and dijets from W-boson 
decay in PbPb collisions as compared with pp interactions are predicted for top anti-top pair 
production~\cite{Bhattacharya:2012gy}. Single and pair top-quark production cross sections in proton-lead and 
lead-lead collisions at the energies of LHC and Future Circular Collider
(FCC) have been estimated recently in~\cite{d'Enterria:2015jna} 
with next-to-leading-order perturbative QCD calculations including nuclear
parton distribution functions.

In our paper we analyze the influence of nuclear effects on single top decay pattern in PbPb 
collisions at nominal LHC energy $\sqrt s_{\rm NN}=5.5$ TeV in the frameworks of 
CompHEP~\cite{Boos:2004kh} and PYQUEN~\cite{Lokhtin:2005px} event generators. The cross section 
calculation of single top production process is done with MCFM~\cite{Campbell:2004t}, subsequent generation of nucleon-nucleon events 
is done with CompHEP, while PYQUEN being used to simulate in-medium rescattering and energy loss 
of top quark partonic decay products. The neutron and proton content in Pb nuclei, and 
nuclear parton distribution functions have been taken into account. The medium-induced 
modification of different characteristics of single top quark decay products (invariant mass 
distribution of W-boson and b-jet from top quark decays, the fraction of top 
quarks observed with associated partner jet, transverse momentum distribution of 
associated jets), and the charge asymmetry of top/anti-top quark yields due to the isospin 
effect are investigated.

\section{Simulation of single top quark production}

\begin{figure}
\resizebox{0.49\textwidth}{!}{\includegraphics{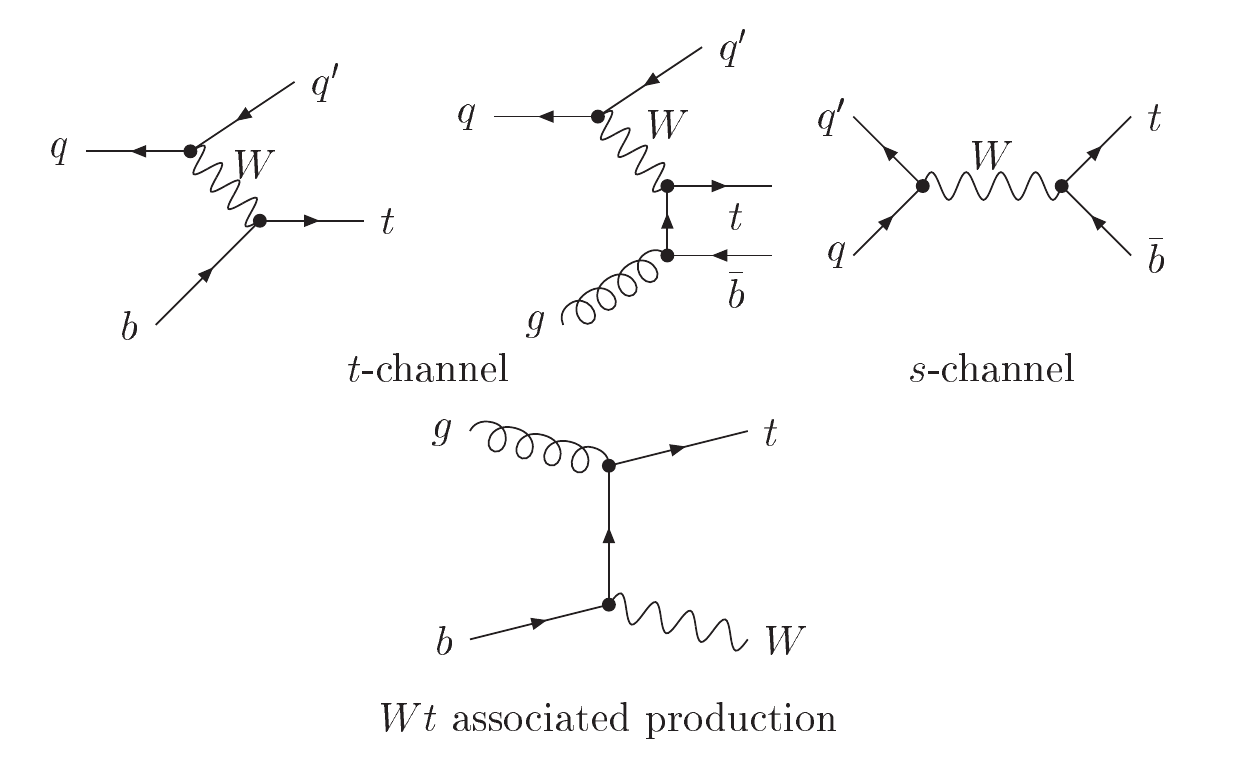}}
\caption{The representative diagrams for the single top production 
at hadron colliders} 
\label{diag_lhc}
\end{figure}

At hadron and lepton colliders top quarks are produced either inpairs or singly. 
The representative diagrams for the single top production are shown in Fig.~\ref{diag_lhc}.

Three mechanisms of the single top production are distinguished 
by the virtuality $Q^2_W$ of the W-boson involved:
$t$--channel ($Q^2_W < 0$),
$s$--channel ($Q^2_W > 0$),
associated tW ($Q^2_W = M^2_W$).
At LHC energies, most of single top quarks are produced with associated partner jet 
in $t$--channel. In our studies we are taking into account only $t$--channel, as the most significant one.

We consider three scenarios: 
(1) without any nuclear effect; 
(2) with initial state effect (nuclear shadowing); 
(3) with both initial and final (jet quenching) state effects.
This notations are used on all the plots and are described in details in this section and Table~\ref{sim_tab}.

\begin {table}
\begin{tabular}{ | l | c | c |}
\hline
       & EPS09 & PYQUEN \\ \hline
(1) pp & - & - \\ \hline
(2) PbPb  & + & - \\ \hline
(3) PbPb + PYQUEN  & + & + \\
\hline
\end{tabular}
\caption{For all scenarios general simulation procedure (CompHEP and MCFM with MSTW2008NNLO + PYTHIA) is implemented. 
Simulation procedure differences are shown in the tabel.}
\label{sim_tab}
\end {table}

\subsection{Initial state}

We use CompHEP for effective NLO generation~\cite{Boos:2006t} of single top events at $\sqrt s_{\rm NN}=5.5$ TeV, 
where exact NLO cross sections are obtained with MCFM~\cite{Campbell:2004t}. 
\mbox{CompHEP} is a general, tree level generator, which allows one to study various SM and 
BSM processes $2\rightarrow N$  (up to $N = 6$) in the framework of the usual technique 
of Feynman diagrams squared for different models. It generates, squares and 
symbolically calculates a set of Feynman diagrams for a given process and creates a numeric 
Monte-Carlo generator for the process. This generator allows one to compute cross sections 
(with applied cuts), to build distributions and to generate events with partons in 
the final state; the initial partons are convoluted with parton distribution 
functions (PDF). In nucleus-nucleus AA collisions, the parton flux is enhanced by the number 
$A^2$ of nucleons in the nucleus and – modulo small (anti)shadowing effects in the nuclear PDF.
In our calculations we have used MSTW2008 PDF~\cite{Martin:2009st} at NNLO with 
nuclear corrections PDF EPS09~\cite{Eskola:2009uj} from LHAPDF library ~\cite{Whalley:2009bg}. 
It allowed to take into account nuclear (anti)shadowing modifications of the bound relative to free 
nucleons as well as the different isospin (u- and d-quark) content of Pb ion given by its different proton (Z=82) 
and neutron (N=126) numbers. The latter effect is important in the case of electroweak 
isospin-sensitive processes like W, Z, prompt photon and single top quark production.

\begin{figure}
\begin{center}
\resizebox{0.5\textwidth}{!}{\includegraphics{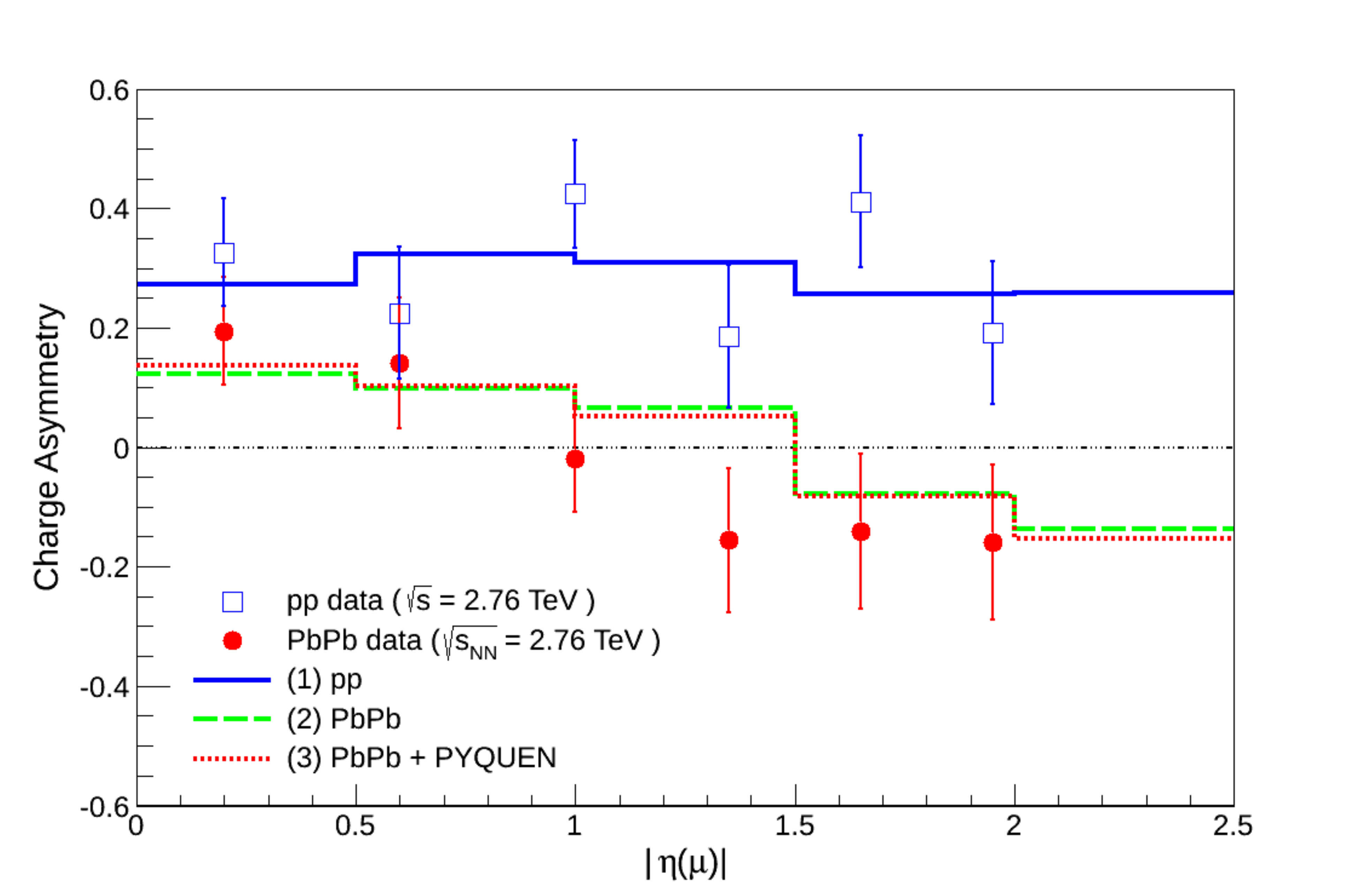}}
\caption{Experimental data for charge asymmetry $({N_{W+} - N_{W-}})/({N_{W+} + N_{W-}})$ is 
shown for pp(blue open squares) and PbPb(red-filled circles) at $\sqrt s =2.76$~\cite{Chatrchyan:2012nt}.
Modeling for pp collisions, first scenario, is shown with blue solid line. Second, PbPb collisions - green dashed line. 
The third one, PbPb collisions with final state effects are shown with red dotted line. Scenario details are described in Tab.~\ref{sim_tab}.}
\label{WW_char_asym}
\end{center}
\end{figure}

The parton showering and hadronization for generated events have been performed with PYTHIA$\_$6.4 event 
generator~\cite{Sjostrand:2006za}. The correctness of our simulation procedure Tab.~\ref{sim_tab} is confirmed 
in a good agreement with the CMS data on W-production cross section and rapidity 
dependence of $W^+,W^-$ charge asymmetry (Fig.~\ref{WW_char_asym}) in PbPb collisions at $\sqrt s_{\rm NN}=2.76$ TeV~\cite{Chatrchyan:2012nt}.

\begin{figure}
\begin{center}
\resizebox{0.5\textwidth}{!}{%
\includegraphics{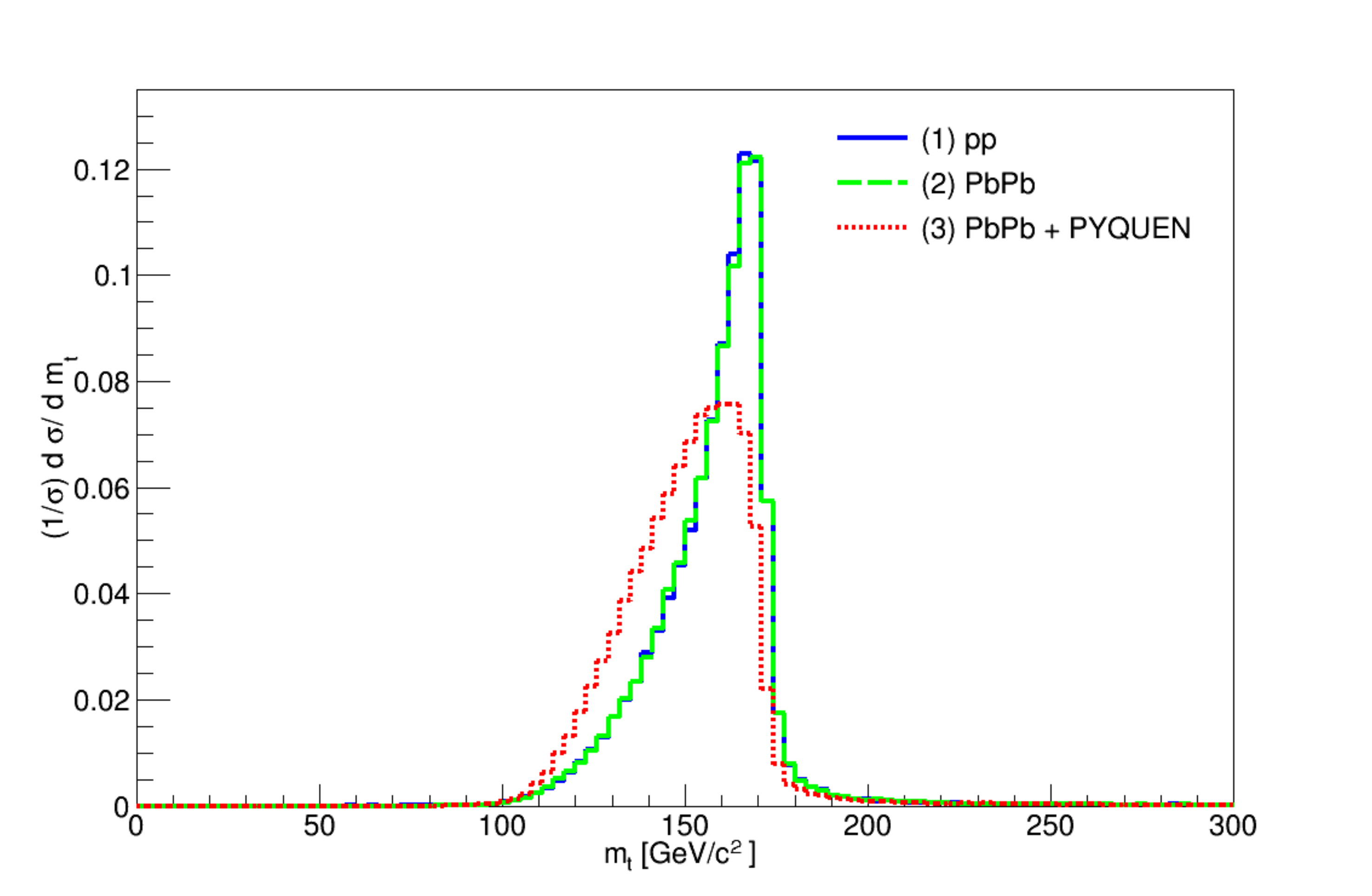}
}
\end{center}
\caption{Distribution of invariant top mass reconstructed from b-jet and W-boson for different scenarios.
Scenario details are described in Tab.~\ref{sim_tab}.} 
\label{m_j_t}
\end{figure}

\begin{figure*}
\centering
\mbox{
\subfigure[]{
\resizebox{0.5\textwidth}{!}{%
\includegraphics{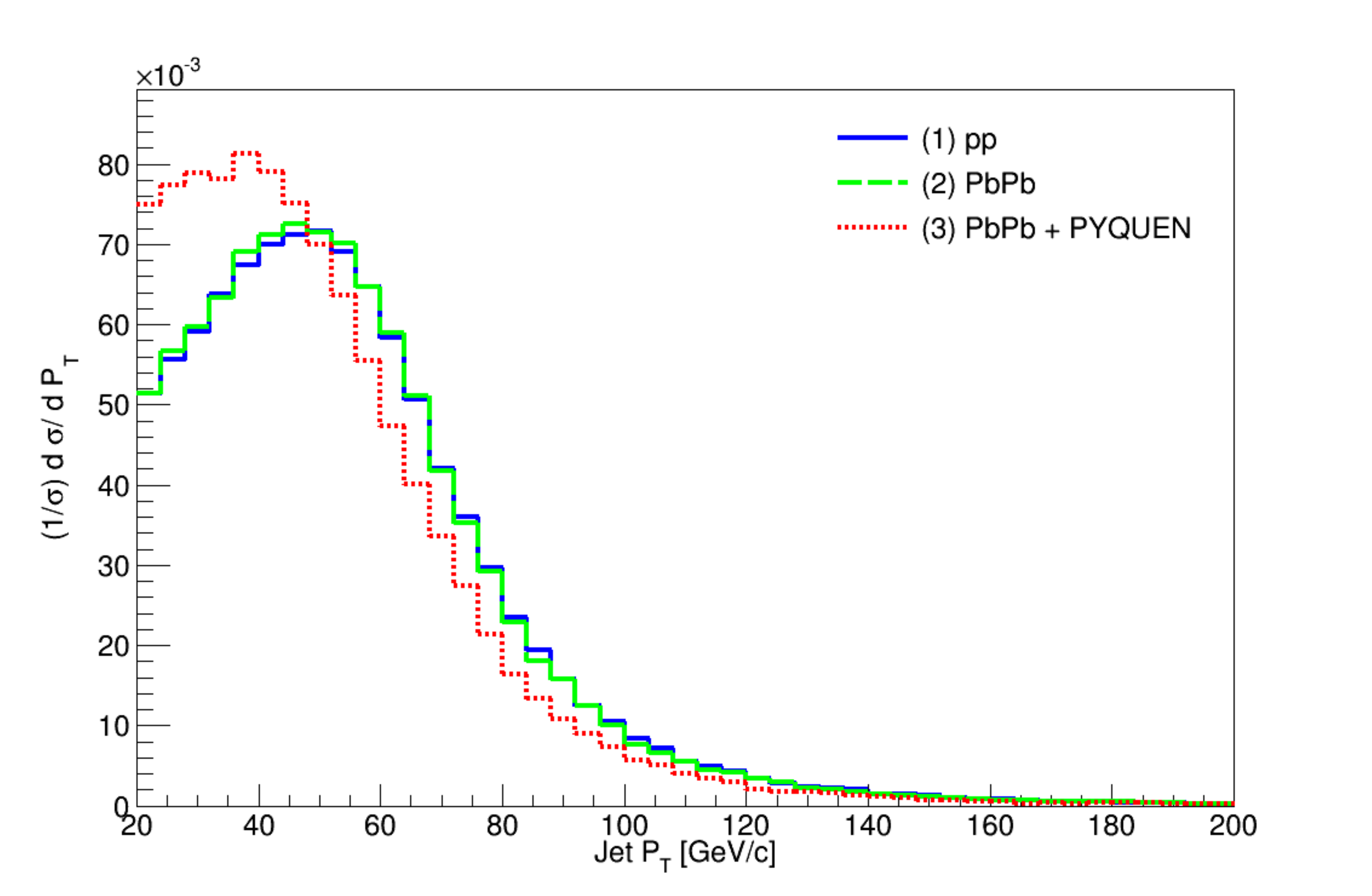}}
} \quad
\subfigure[]{
\resizebox{0.5\textwidth}{!}{%
\includegraphics{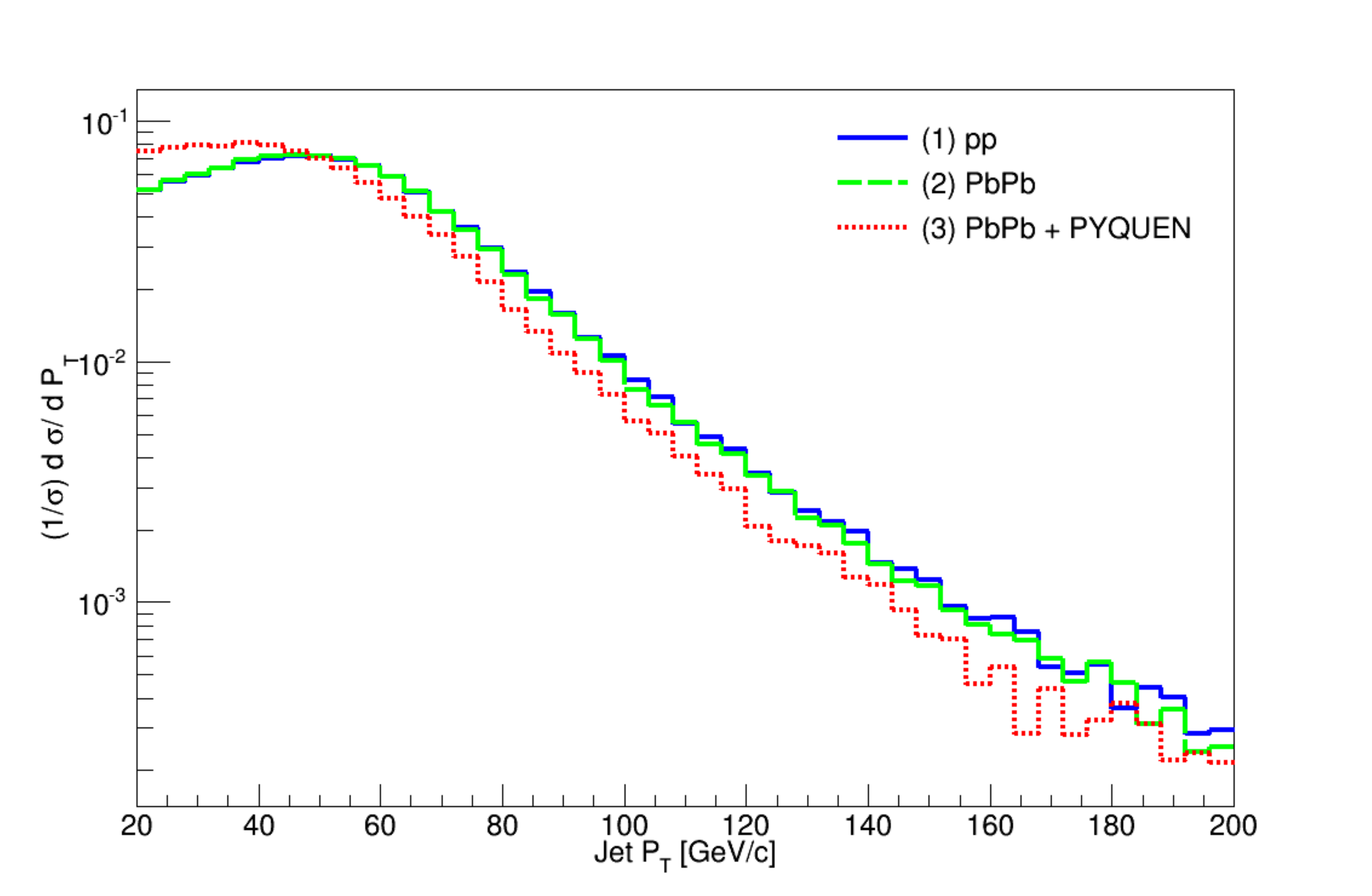}}
}
}
\caption{Distribution of transverse momentum of b-jet from top quark decay for different scenarios. 
Left figure is in linear scale right in logarithmic.
Scenario details are described in Tab.~\ref{sim_tab}.} 
\label{jet_b_t_pt}
\end{figure*}

\begin{figure*}
\centering
\mbox{
\subfigure[]{
\resizebox{0.5\textwidth}{!}{%
\includegraphics{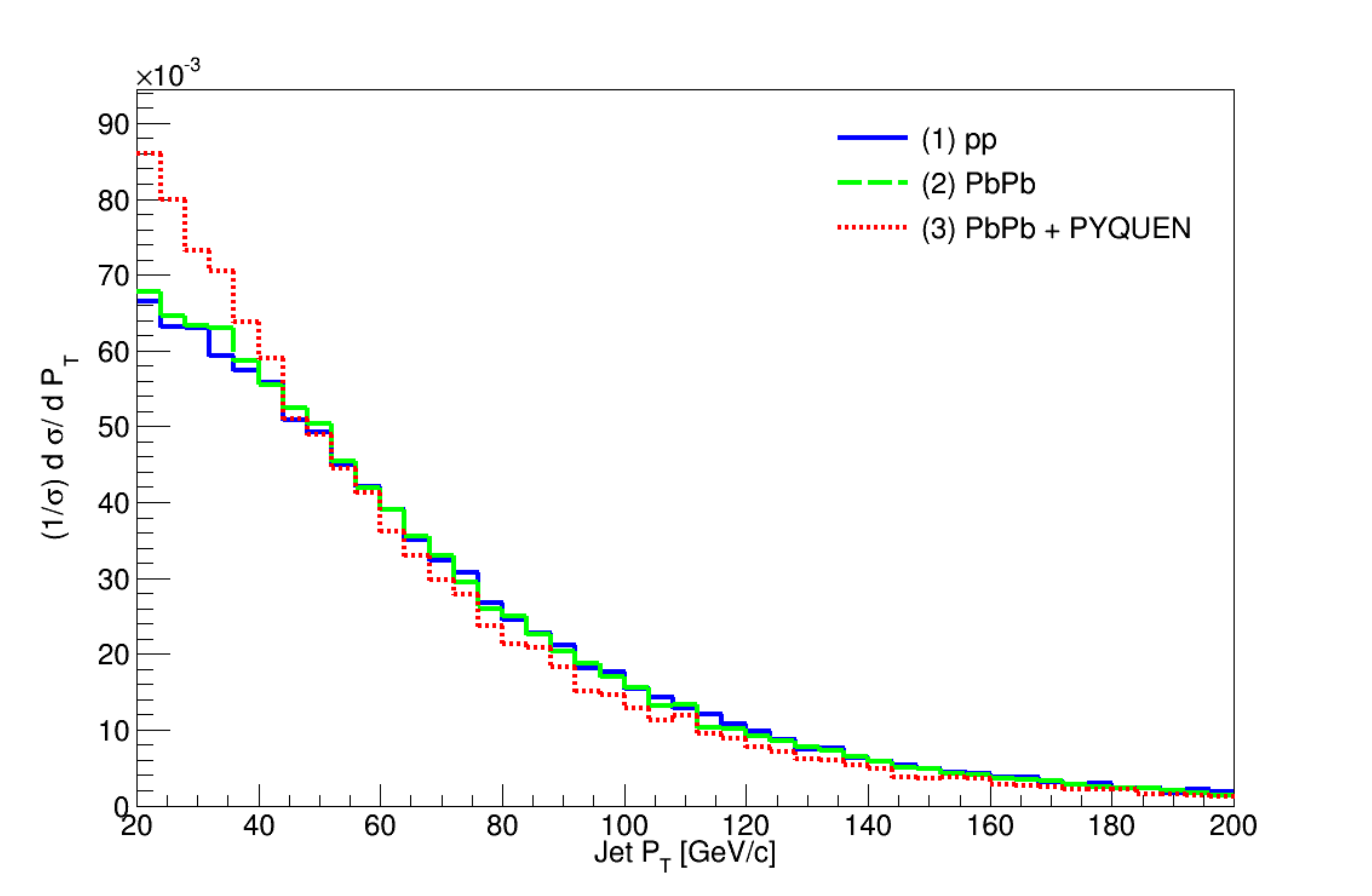}}
} \quad
\subfigure[]{
\resizebox{0.5\textwidth}{!}{%
\includegraphics{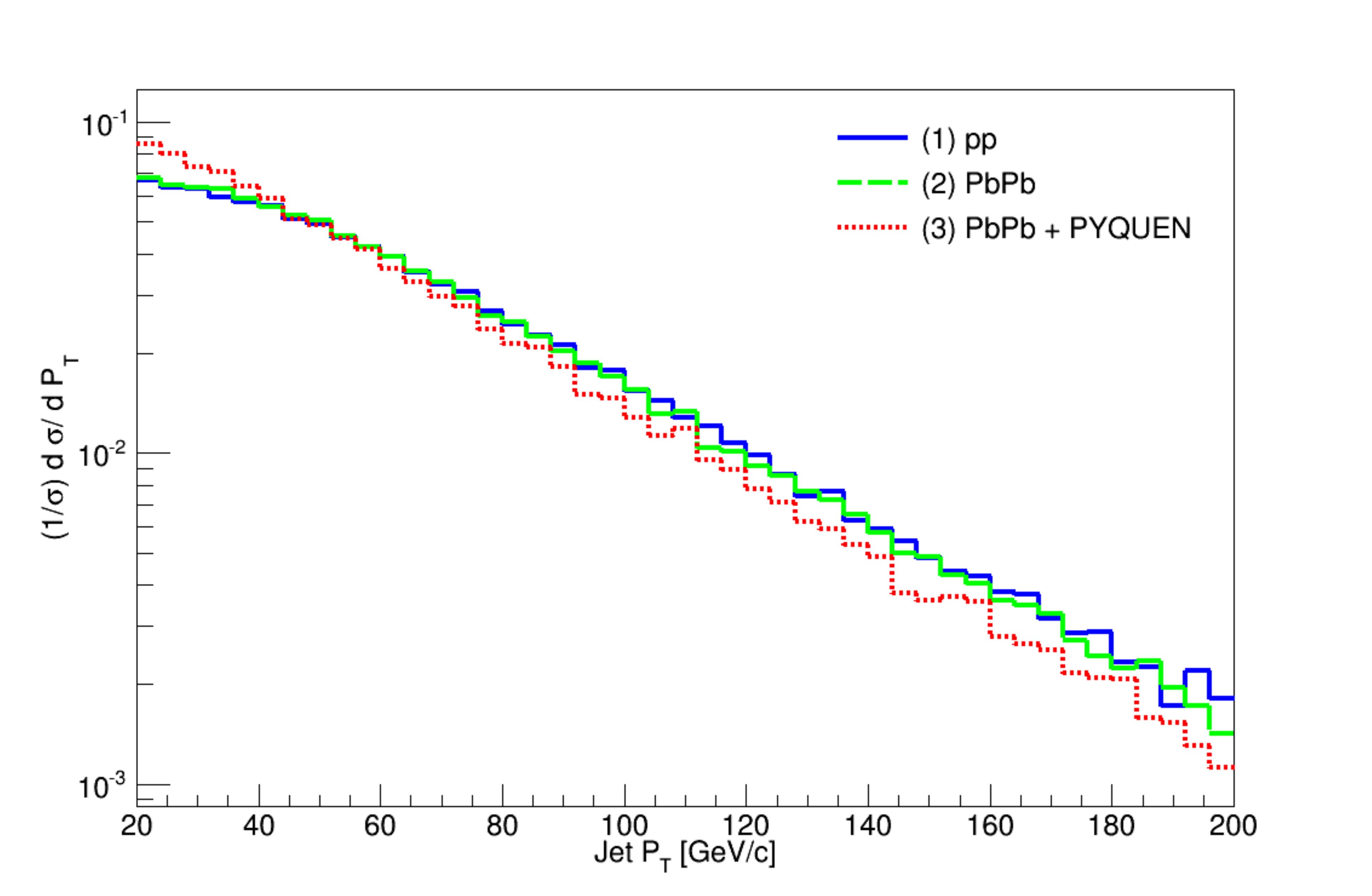}}
}
}
\caption{Distribution of transverse momentum of jet from light quark for different scenarios.
Left figure is in linear scale right in logarithmic.
Scenario details are described in Tab.~\ref{sim_tab}.} 
\label{jet_l_q_pt}
\end{figure*}

\begin{figure}
\resizebox{0.5\textwidth}{!}{%
\includegraphics{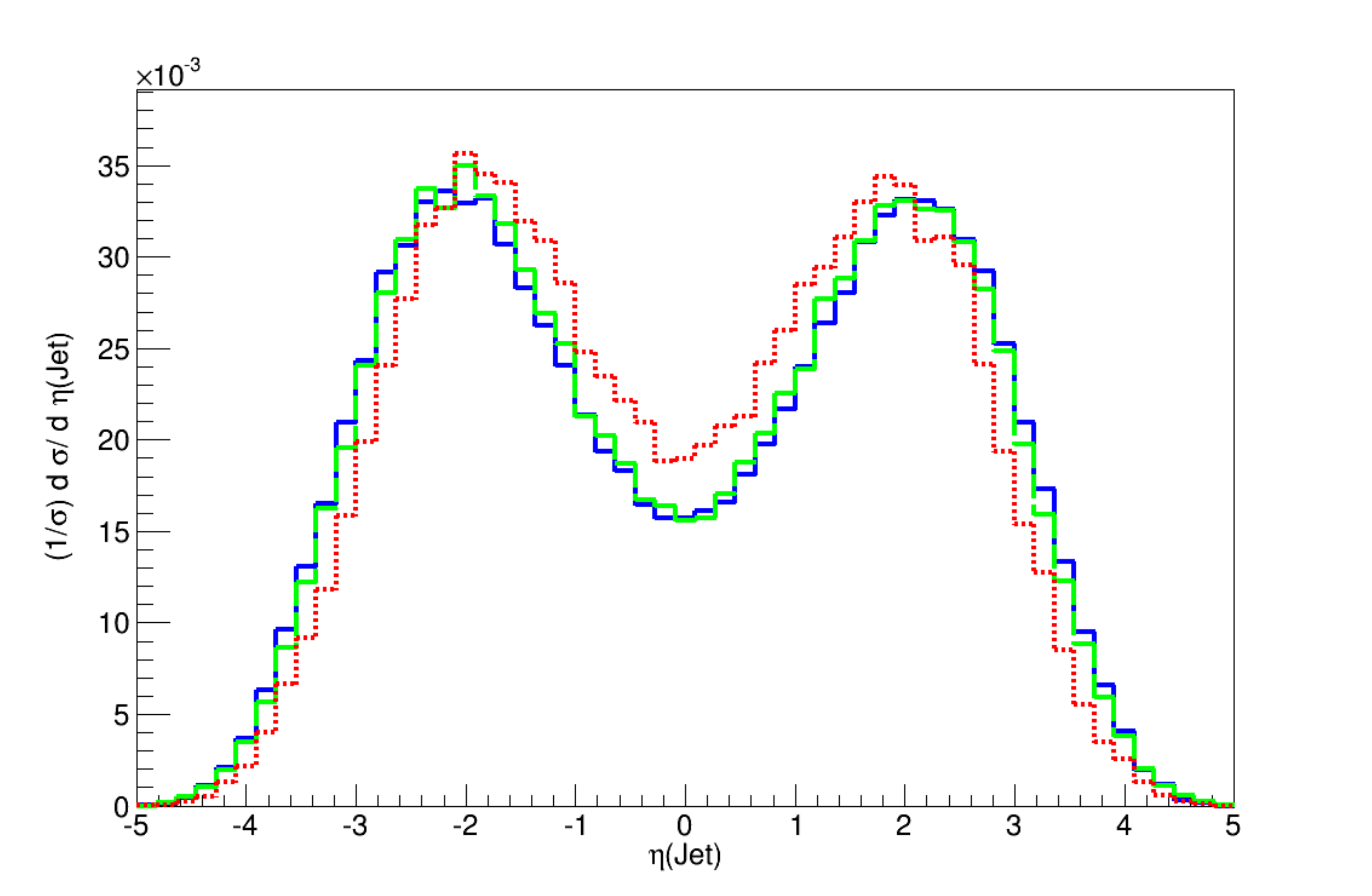}
}
\caption{Distribution of pseudorapidity of jet from light quark.
Scenario notations are similar to other figures, details are described in Tab.~\ref{sim_tab}.} 
\label{jet_q_eta}
\end{figure}

We consider the channel with leptonic ($t \rightarrow Wb \rightarrow b \nu_l$, $l=\mu, e$) top 
quark decay modes. The possibility to reconstruct W-bosons in PbPb collisions with CMS detector 
using missing-$E_T$ technique has been revealed in~\cite{Chatrchyan:2012nt}. The efficiency of 
b-tagging in PbPb collisions is estimated on the level of $\sim 50 \%$~\cite{Chatrchyan:2013exa}.
Thus the reconstruction of top quark in PbPb collisions in the leptonic decay mode looks quite 
realistic. The reconstruction of hadronically decayed single top quarks ($t \rightarrow Wb 
\rightarrow bq\bar{q}'$) seems practically impossible even in pp collisions due to huge QCD 
background. 

In order to estimate the expected event rate for realistic geometrical acceptance and kinematic 
cuts, the selection requires pseudorapidity coverage $\mid \eta \mid < 5$ for light-flavor quark jets, presence of at least one b-tagged jet
with $\mid \eta \mid < 2.5$  and $\mid \eta \mid < 2.5$ for 
leptons (muons and electrons), where $\eta =  -\ln [ \tan(\theta/2)]$. The criteria corresponds to the conditions of ATLAS and 
CMS experiments. The cuts on transverse momenta of leptons $p_{\rm T}^{\rm l}> 20$ GeV/$c$ and jet transverse energy $E_{\rm T}^{\rm jet}>20$ GeV were applied. 

For simplicity jet clustering is done with PYCELL Pythia routine. Cone based criteria 
$\Delta R<0.4$, where  $\Delta R = \sqrt{\Delta {\eta}^2  + \Delta {\varphi}^2 }$,
is used to associate jet to parton. In single top quark production topology presupposes 
two hard jets with high $P_T$, b-jet from top quark decay and light quark jet and one soft jet with low $P_T$ from additional b-quark.

We estimate efficiency of kinematic acceptance as $\frac{N_{pass}}{N_{total}}$, where 
$N_{pass}$ is number of events passed kinematic cuts, $N_{total}$ is full number of events generated by Monte-Carlo, and it is $\sim65\%$.
B-tagging acceptance is estimated as $$\frac{\epsilon \times N_{1 b jet} + {(1-(1-\epsilon)}^2) \times N_{2 b jet} }{N_{1 b jet} + N_{2 b jet}},$$
where $N_{1 b jet}, N_{2 b jet}$ are number of events passed kinematic cuts with reconstructed one and two b-jets correspondingly, $\epsilon$ is 
b-tagging efficiency. For $\epsilon=0.5$ we can estimate acceptance $\sim52\%$. Total acceptance as $\sim31-35\%.$

Single top t-channel production NLO cross section is calculated with MCFM and is $38.6$ pb. 
We assume equal branching $\frac{1}{9}$ for muon and electron and total acceptance as $35\%$. Then the estimated pp cross section for single top quark production in leptonic 
mode (electron and muon channels without any contribution from tau-channel) satisfying above cuts is $\sim 3$ pb, and PbPb cross section is estimated as $\sim ~ 0.13 \rm \mu b$. 
Our numbers are compatible with ones obtained in~\cite{d'Enterria:2015jna} (taking into account that different nucleon PDF and little different kinematic cuts were used). 
The corresponding event rate in a one month PbPb run ($T=10^6$ seconds), with nominal luminosity 
$L = 10^{27}~$cm$^{-2}$s$^{-1}$, is estimated as $N_{\rm ev}= T \sigma_{\rm PbPb} L \sim 130$ 
(this corresponds to $1 \rm nb^{-1}$ integrated luminosity data). 
In the case of increasing LHC luminosity for lead beams in the future (upgrade plans for HL-LHC) 
by a factor of $10$, the number of detected events with single top quark production may increase up to  $\sim1300$.

The main background sources for single top production in pp collisions
are W+jets, $t \bar{t}$ pairs and QCD jets~[\cite{Chatrchyan:2012tc},~\cite{G.Aad:2012}]. It is natural to expect that
these processes will still form the main background for single top
production in heavy ion collisions. However their significance can be
modified, as well as influence of initial and final state nuclear effects
on them may differ. Background modelling and estimation of signal statistical
significance is complex and important task, which however goes beyond the
current paper topic. It could become the subject of our future studies.

\subsection{Medium-induced final state interactions}
\label{sec:pyq}

After specifying initial partonic state from top quark decays with CompHEP and PYTHIA,
event-by-event Monte-Carlo simulation of medium-induced rescattering and energy loss of jet
partons is performed with PYQUEN event generator~\cite{Lokhtin:2005px}. This model is 
constructed as a modification of an event obtained with the PYTHIA. The approach 
describing the multiple scattering of hard partons is based on accumulated energy loss via gluon 
radiation which is associated with each parton scattering in expanding quark-gluon fluid. It 
also includes the interference effect in gluon emission with a finite formation time using the 
modified radiation spectrum as a function of the decreasing temperature. The model takes into 
account radiative and collisional energy loss of hard partons in longitudinally expanding 
quark-gluon fluid, as well as the realistic nuclear geometry. The event-by-event simulation 
procedure in PYQUEN includes the generation of initial parton spectra with PYTHIA and production 
vertexes at the given impact parameter of AA collision; the rescattering-by-rescattering 
simulation of the parton path in a dense nuclear overlapping zone inducing its radiative and 
collisional energy loss; and the final hadronization according to the Lund string model for hard 
partons and in-medium emitted gluons.

The strength of partonic energy loss in PYQUEN is determined mainly by the initial maximal 
temperature $T_0^{\rm max}$ of hot fireball in central PbPb collisions, which is achieved 
in the center of nuclear overlapping area at mid-rapidity. The transverse energy density in 
each point inside the nuclear overlapping zone is supposed to be proportional to the 
impact-parameter dependent product of two nuclear thickness functions $T_A$ in this point,  
$\varepsilon (r_1, r_2) \propto T_A(r_1) T_A(r_2)$ (here $r_{1,2}$ are the transverse distances 
between the centres of colliding nuclei and the parton production vertex). The rapidity 
dependent spreading of the initial energy density around mid-rapidity $y=0$ is taken in the 
Gaussian-like form. The radiative energy loss in the model roughly are proportional to $T_0^3$, 
and  collisional loss roughly are proportional to $T_0^2$. Such strong dependence of jet quehcning 
effect on an initial temperature allows us to fix this model parameter within the experimental 
constraints on jet quenching data. The partonic energy loss in the model depends also on the proper 
time $\tau_0$ of matter formation and the number $N_f$ of active flavors in the medium. The 
variation of $\tau_0$ value within its reasonable range has rather moderate influence on the 
strength of partonic energy loss. The jet quenching gets stronger at larger $\tau_0$ due to 
slower medium cooling, which implies the jet partons spending more time in the hottest regions, 
and as a result the rescattering intensity gets stronger~\cite{Lokhtin:2011qq}. Increasing $N_f$ 
result in larger energy loss due to the extension of medium density in this case. The parameter values 
$T_0^{\rm max}=1.1$ GeV, $\tau_0=0.1$ fm$/c$ and $N_{\rm f}=0$  (gluon-dominated plasma), and also 
the ``wide-angle'' pa\-ra\-met\-ri\-za\-ti\-on of medium-induced gluon radiation were used for 
current investigation. PYQUEN with such settings (just with $T_0^{\rm max}=1$ GeV) reproduces the LHC 
data on the dijet asymmetry~\cite{Lokhtin:2011qq}, charged particle nuclear modification factor 
(up to $p_{\rm T} \sim 100$ GeV/$c$)~\cite{Lokhtin:2012re}, jet fragmentation function and jet 
shapes~\cite{Lokhtin:2014vda} in PbPb collisions at $\sqrt{s_{\rm NN}}=2.76$~TeV.

\section{Results}
\label{sec:results}

Figure~\ref{m_j_t} shows the invariant mass distributions of W-boson and b-jet from top quark decays
in minimum bias (integrated over all impact parameters) PbPb interactions and pp collisions at 
$\sqrt s_{\rm NN}=5.5$ TeV. The smearing and decreasing mean and maximum values of this distribution 
from top quark mass are shown in 
PbPb events relatively to pp collisions. Transverse momentum spectra of jets associated with top
quark is presented in Figure~\ref{jet_b_t_pt}. Significant softening of this distribution and peak shifting  
from the half mass of W-boson is seen for PbPb collisions. Softening of jet from light quark is also observed 
in Figure~\ref{jet_l_q_pt} and its pseudorapidity modification is seen on Figure~\ref{jet_q_eta}.

\begin{figure}[]
      \resizebox{0.49\textwidth}{!}{\includegraphics{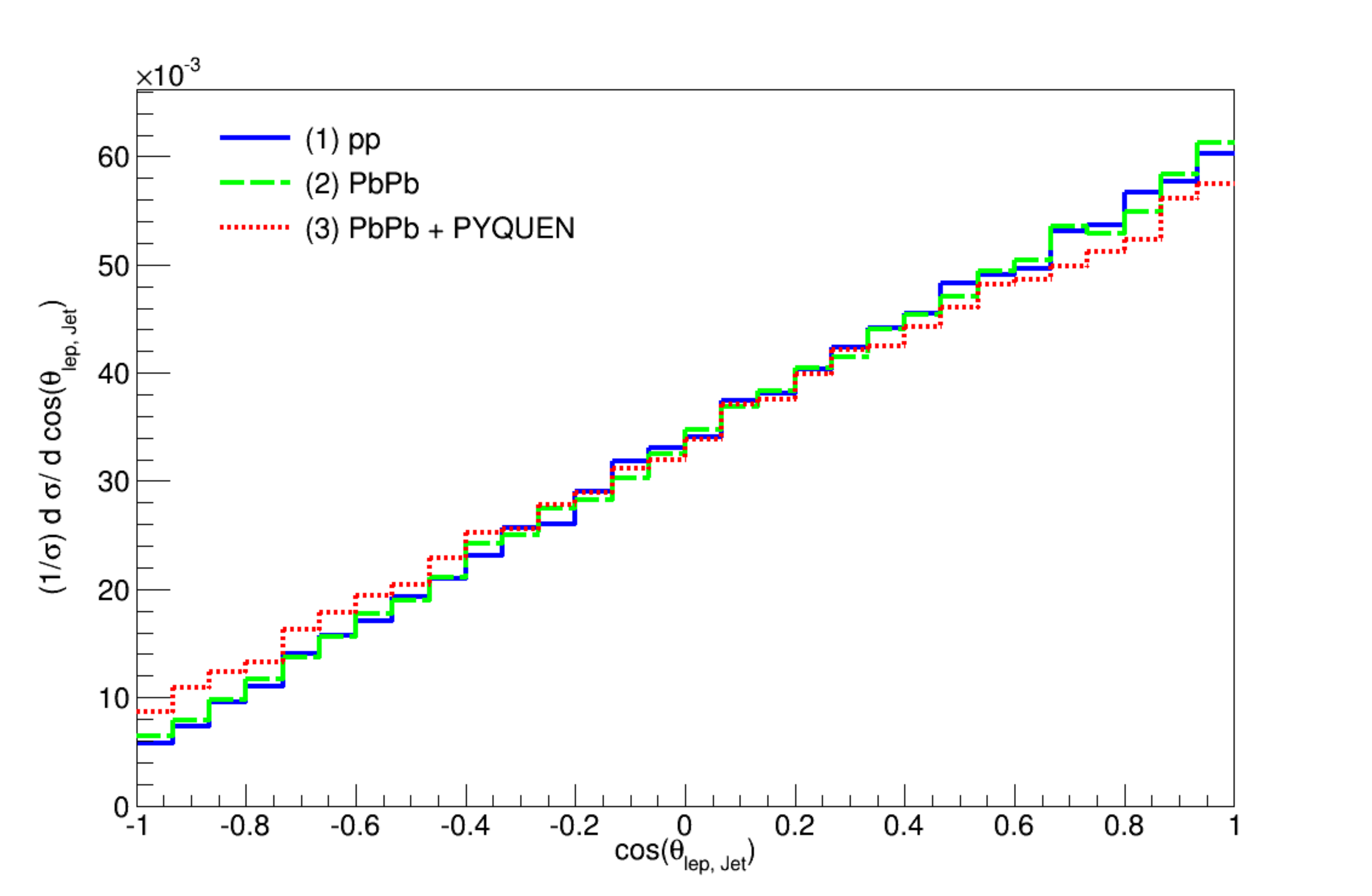}}
\caption{Cosine of angle between lepton from top decay and jet from light 
quark associated with the top in top quark rest frame. Scenario details are described in Tab.~\ref{sim_tab}.} 
\label{cos_lj}
\end{figure}

One expects that any lepton kinematic distributions are not practically
affected by the nuclear medium. However, the distribution on the angle
taken in the top rest frame between the lepton from the top decay and the
momentum of the light jet produced in association with the top is
slightly affected due to jet quenching of all the jets not only from the
top decay. The distribution on this important variable for separating the
single top signal from backgrounds is shown in the Figure~\ref{cos_lj}. 

\begin{figure}[t]
      \resizebox{0.49\textwidth}{!}{\includegraphics{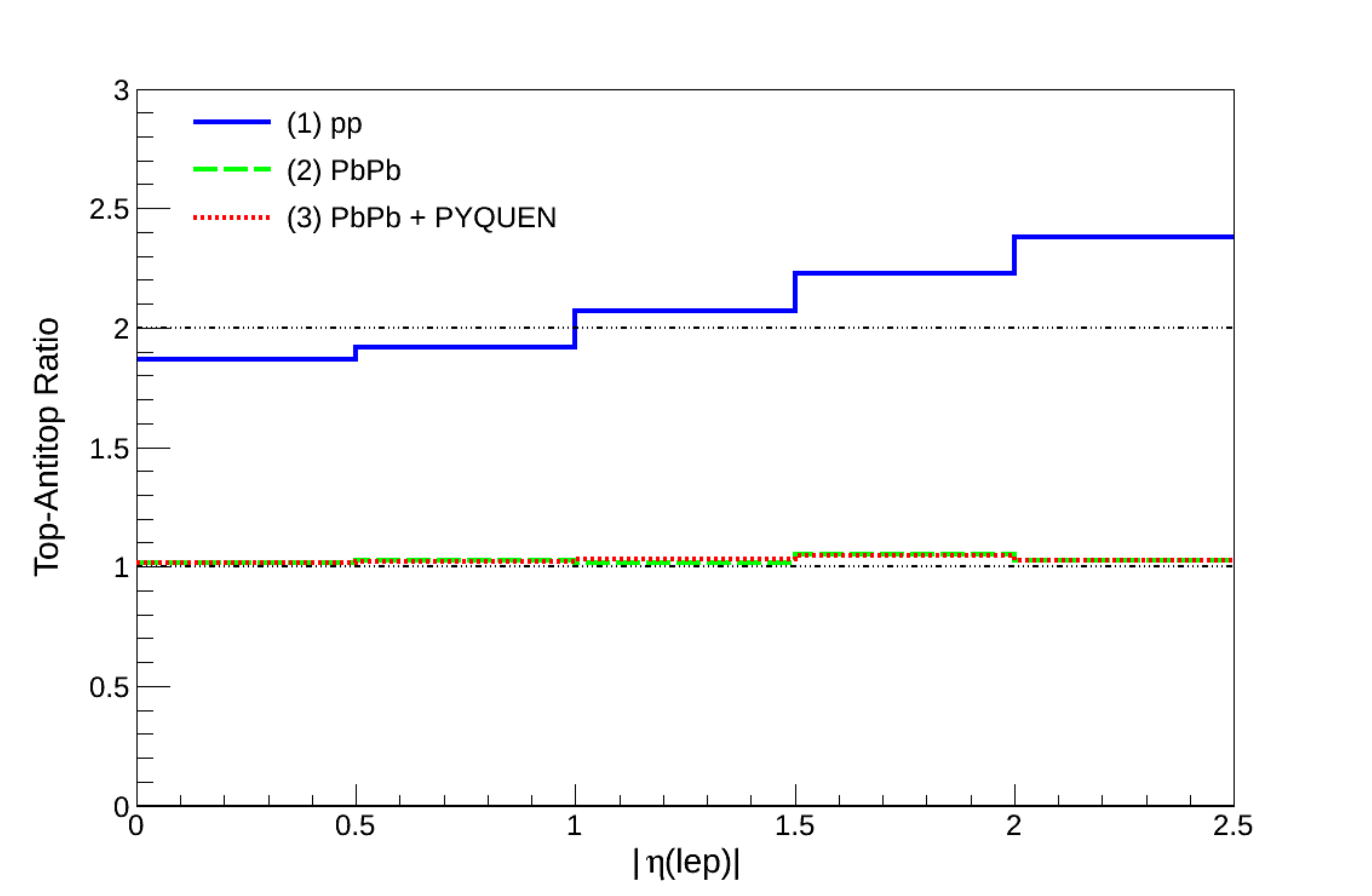}}
\caption{Top/anti-top ratio $N_{t} / N_{\bar{t}}$. Scenario details are described in Tab.~\ref{sim_tab}.} 
\label{tT_ratio_asym}
\end{figure}

\begin{figure}[]
      \resizebox{0.49\textwidth}{!}{\includegraphics{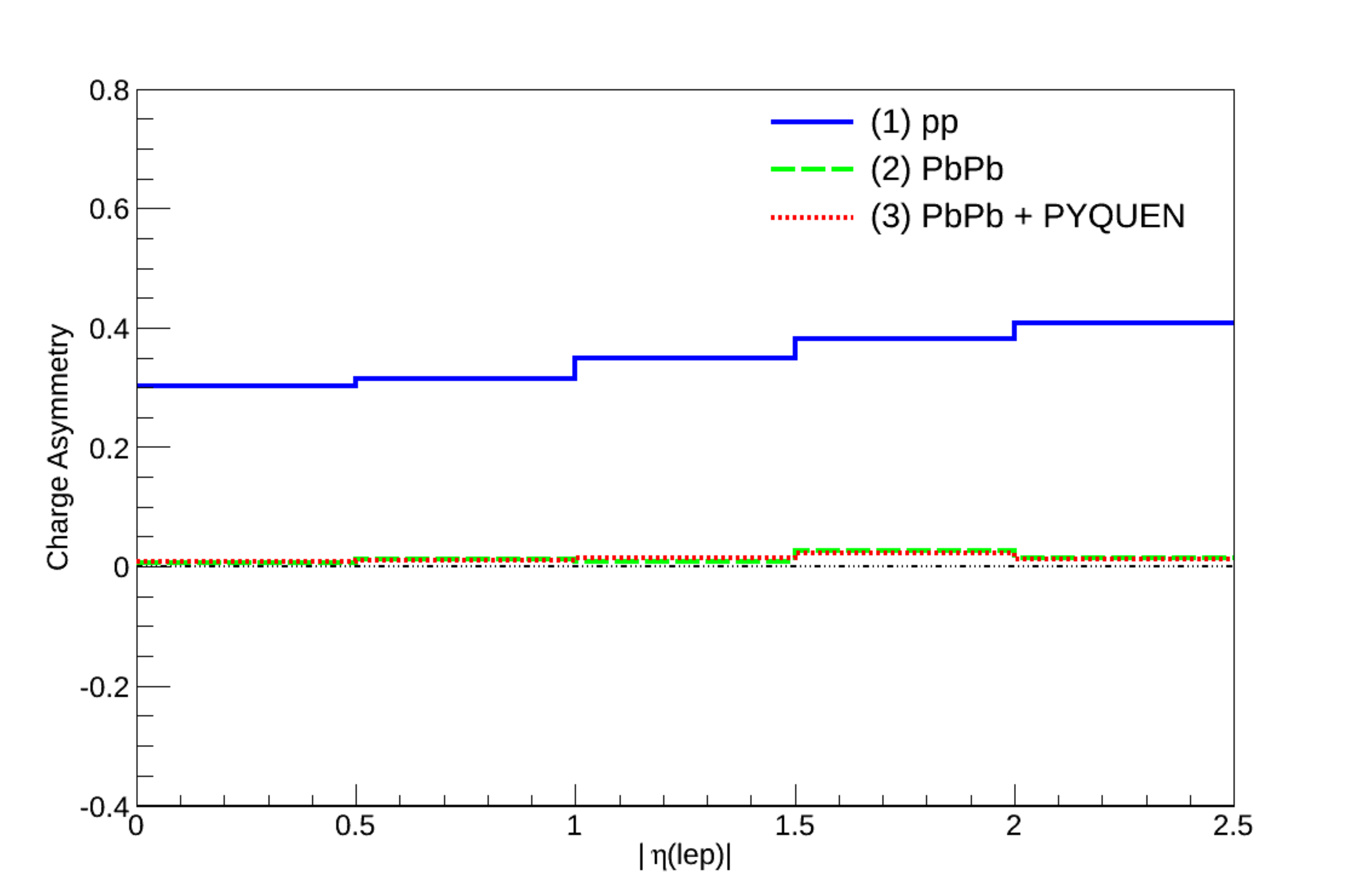}}
\caption{Charge asymmetry $({N_{t} - N_{\bar{t}}})/({N_{t} + N_{\bar{t}}})$ for three scenarios. 
Scenario details are described in Tab.~\ref{sim_tab}.}
\label{tT_charge_asym}
\end{figure}

Another interesting (initial state) nuclear effect, which is not directly related to quark-gluon
matter formation but reflects the neutron and proton content in Pb nuclei, is significant decrease in
ratio of top to anti-top quarks cross sections in PbPb with respect to pp collisions at the same
beam energy per nucleon pair. The predicted ratios of $t/\bar{t}$ yields for minimum bias PbPb 
events and pp collisions at $\sqrt s_{\rm NN}=5.5$ TeV are $\sim1$ and $\sim2$  respectively. Dependence 
of these ratios from lepton $\eta$ is shown on the Figure~\ref{tT_ratio_asym}. The same effect leads to the 
disappearance of charge asymmetry in PbPb collisions (Fig.~\ref{tT_charge_asym}).

\section{Conclusions}
\label{sec:summ}

The t-channel single top production in PbPb collisions at nominal LHC energy 
$\sqrt s_{\rm NN}=5.5$ TeV has been analyzed in the frameworks of MCFM(calculation of cross
sections), CompHEP (generation of events), PYTHIA (parton showering and hadronization) and PYQUEN (medium-induced  partonic
rescattering and energy loss) Monte-Carlo models. The neutron and proton content in Pb nuclei, 
and nuclear parton distribution functions have been taken into account. Single top quark production has 
a large enough cross section and visible event rate for the nominal LHC luminosities and is open to study in PbPb
collisions.

Medium-induced partonic energy loss can result in the modification of different 
characteristics of single top quark decay products in PbPb collisions as compared with pp
interactions. We predict smearing and decreasing mean and maximum values of the invariant mass 
distribution of W-boson and b-jet from top quark decays, significant softening 
the transverse momentum spectrum of jets associated with top quark and shifting its peak from 
the half mass of W-boson. 

Distribution on the angle taken in the top rest frame between the 
lepton from the top decay and the momentum of jet from light quark produced in association with the top is only
slightly affected and may be used for separation single top signal from backgrounds in future \mbox{analysis}.

Besides due to the nuclear isospin effect, the ratio of top to anti-top
quark cross sections becomes close to unity in PbPb being by a factor
$\sim 2$ smaller than in pp collisions.

\begin{acknowledgments}
We thank our colleagues from CMS collaboration for fruitful cooperation. 
This work was supported by Russian Foundation for Basic Research (grants 
12-02-91505, 13-02-01050) and Grant of President of Russian Federation for
Scientific Schools Supporting No. 3042.2014.2.
\end{acknowledgments}

\end{document}